\documentclass[showpacs,twocolumn,prb]{revtex4-1}
\usepackage{graphicx}
\graphicspath{{figures/}} 
\usepackage{amsmath}
\usepackage{enumitem}
\usepackage{amsfonts}
\usepackage{amssymb}
\usepackage[caption=false]{subfig}

\newcommand{\bwt}{\begin{widetext}}
\newcommand{\ewt}{\end{widetext}}

\newcommand{\be}{\begin{equation}}
\newcommand{\ee}{\end{equation}}
\def\bea {\begin{eqnarray}}
\def\eea {\end{eqnarray}}

\usepackage{hyperref}
\hypersetup{
    colorlinks=true,
    linkcolor=blue,
    filecolor=magenta,      
    urlcolor=blue,
    citecolor=red,
}

\usepackage{color}

\usepackage{comment}
\def\comment#1{}

\usepackage{ulem}   

\begin{document}

\title{The Modified Airy Function Approximation Applied to the Double-Well Potential}

\author{N. Wine, J. Achtymichuk and F. Marsiglio}
\affiliation{Department of Physics, University of Alberta, Edmonton, AB, Canada T6G~2E1}

\begin{abstract}
The single harmonic oscillator and double-well potentials are important systems in quantum mechanics. The single harmonic oscillator is {\it the} paradigm in physics, and
is taught in nearly all beginner undergraduate classes, while the double-well potential illustrates the two important principles of quantum tunnelling and linear superposition. 
While exact analytical solutions of the Schr\"odinger equation exist for both of these potentials, they are also employed to benchmark the use of approximate techniques which
may be the only recourse for more complicated potentials.
In this paper, we review the Wentzel-Kramers-Brillouin (WKB) approximation for both these potentials. While this approximation is known for its accurate energies,
we will instead emphasize how poor the WKB wave functions are.  The inaccuracy of the WKB wave functions will then motivate us to adopt the lesser-known 
Modified Airy Function (MAF) approximation, which alleviates
the deficiencies of the WKB wave functions. We will review the MAF solution to the simple harmonic oscillator potential, and then apply the MAF to the double-well
potential. We find accurate eigenvalues and, more importantly, very accurate wave functions. We conclude with the suggestion that an introduction to the MAF should
be included in undergraduate courses to complement the WKB.
\end{abstract}

\date{\today }
\maketitle

\section{Introduction} \label{sec: intro}

The two key potentials in quantum mechanics are arguably the simple harmonic oscillator potential (SHO) and the double-well potential (DWP).
The simple harmonic oscillator is used to understand the low energy behaviour of many systems of interest in various branches of physics, while the double-well potential
represents the paradigm for so-called Schr\"odinger Cat states. Both of these can be studied in quantum mechanics with the time-independent Schr\"odinger equation (TISE),
and both have exact analytical solutions. For the SHO, these solutions consist of Hermite polynomials,\cite{Griffiths_3rd_edition} which are elementary functions in the sense that they are polynomials, although they usually do not appear as a button on a calculator.
For the form (to be described below) of the DWP that we focus on in this paper, the solutions can be determined in terms of non-elementary functions known as confluent hypergeometric functions.\cite{merzbacher98} 

While the eigenstates for the TISE are analytically known for these potentials, it is often the case that more realistic potentials are required for the problem
at hand.
In this case, a numerical or approximate solution to the TISE is required. 
One approximation method that is often used, for example, is the Wentzel-Kramers-Brillouin (WKB) approximation, famous historically for understanding alpha decay
and for calculating tunnelling probabilities.\cite{Griffiths_3rd_edition,wiesendanger94} The WKB approximation is usually discussed in undergraduate quantum texts,\cite{Griffiths_3rd_edition} and has been
analyzed in great detail for the double-well potential in Ref.~\onlinecite{Double-potential-well}, where further references and discussion are included. Typically the
simple harmonic oscillator problem is used as an example in textbooks, where the WKB approximation produces exact eigenvalues, and even for the double-well potential,
the accuracy of the eigenvalues is generally
good. Since the WKB relies on a semi-classical approximation, it is generally the case that the WKB energies are 
increasingly accurate for higher energy levels.

While the energy eigenvalues estimated by the WKB tend to be very accurate, the resulting wave function is seldomly critically discussed. For this reason, we revisit
this problem. As we shall see, the WKB wave functions are quite poor for low quantum number. While they improve with increasing quantum number, they continue
to retain discontinuities, which is an unacceptable situation.

In order to improve this situation, we will explain and utilize the Modified Airy Function (MAF) method, first suggested long ago by Langer,\cite{langer35, langer37a} and 
more recently highlighted by Ghatak, Gallawa and Goyal.\cite{MAF_and_WKB_solns,Approx_sol_to_wave_eqn} The first of these latter two references is a monograph 
and provides an introduction to the methodology and further references. To make this article self-contained, we also will explain the methodology for the MAF and apply it 
to both the SHO and DWP problems. 

The philosophy behind this approach is to eliminate the discontinuities that appear in the WKB wave function. The presence of these discontinuities is
generally not appreciated; they arise because the patching between the WKB and Airy wave functions is generally not possible without a 
discontinuity, as our calculations below will show. Note that the discontinuities we refer to are {\bf not} the divergences at the turning points --- these have been circumvented 
by the patching Airy functions. The MAF avoids the so-called patching region altogether, so no discontinuities arise in this method, and the MAF wave function is actually
quite accurate, even for the ground state. As with the WKB, the accuracy increases with quantum number. On the other hand, the eigenvalues are no longer exact for the
SHO. We do not view this as a failure. Rather, it is a more realistic inaccuracy associated with the approximation, and consistent with the error in the wave function.

The price one has to pay to implement the MAF approximation is familiarity with Airy functions. However, it is increasingly the norm that students can call upon various properties of
non-elementary functions (such as zeros of the Airy functions) with various software tools. Similarly, we will generate exact solutions of the TISE numerically, as described in
Refs.~\onlinecite{Matrix_mechanics,Double-potential-well}. The reason is that this study is meant to be generalizable to potentials that do not necessarily have
known analytical solutions, meaning a simple and general-purpose numerical method is required.

In this paper, we will focus on the wave function solutions for both the WKB and MAF approximations, with the goal of convincing the reader that the WKB wave functions
are unacceptable, and that the MAF wave functions
are vastly superior. We will begin with the more familiar simple harmonic oscillator, with many of the details relegated to the Appendix. We will then discuss the approximate
solutions for the double-well potential, as a number of nuances arise in this problem. Much of the WKB work will be taken from Ref.~\onlinecite{Double-potential-well}, except
the wave functions will be entirely new here. We will work through the MAF approximation for this potential in some detail, as this solution has not appeared before. We conclude
with the result that the WKB approximation does not generate accurate (or even sensible) wave functions, whereas the MAF approximation is quite accurate.

\section{The Simple Harmonic Oscillator} 
\label{sec:sho}

As mentioned in the Introduction, the simple harmonic oscillator is perhaps the single most important potential in quantum mechanics. It describes a particle with
mass $m$ trapped in a quadratic potential with spring constant $k \equiv m\omega^2$, where $\omega$ is the characteristic frequency of the potential. The expression
of the potential as a function of position $x$ is

\begin{equation}
    V(x) = \frac{1}{2}m \omega^2 x^2.
    \label{sho_pot}
\end{equation}

The exact solution to the TISE for this potential can be determined by matrix mechanics,\cite{Matrix_mechanics} or are given analytically by: \cite{Griffiths_3rd_edition,townsend}
\begin{eqnarray}
    \psi_n(x) &= &\frac{1}{\sqrt{2^n n!}} \left(\frac{1}{\pi x_{HO}^2}\right)^{1/4} e^{-\frac{1}{2}{\left(\frac{x}{x_{HO}}\right)}^2} H_n\left(\frac{x}{x_{HO}}\right),\\ 
    \nonumber \\
     E_n &= &\hbar \omega(n+{1 \over 2}) \text{ where } n=0,1,2 \dots
     \label{sho_ex_solution}
\end{eqnarray}
where $E_n$ are the eigenvalues, $\psi_n(x)$ are the eigenfunctions, and $x_{HO} \equiv \sqrt{\hbar \over m\omega}$ is the characteristic
length for the harmonic oscillator potential.
The functions $H_n(x)$ are the Hermite polynomials as given, for example, in Refs.~\onlinecite{Griffiths_3rd_edition}~or~\onlinecite{arfken85} ($H_0(y) = 1$,
$H_1(y) = 2y$, $H_2(y) = 4y^2-2$, etc.).\\

\subsection{The WKB Approximation} \label{subsec: SHO WKB}

The details of the WKB approximation are given in Appendix \ref{appsec:wkb}. Here we simply quote the results. 
As mentioned in the Introduction, the eigenvalues turn out to be identical to the exact ones, $E_n^{\rm WKB} = \hbar \omega(n+{1 \over 2}) \text{ where } n=0,1,2 \dots$.
The wave function is given in terms of dimensionless units, defined as
\begin{equation}
    z  \equiv  \frac{x}{x_{\text{HO}}}, \ \     z_t  \equiv  {x_t \over x_{HO}} \  = \  \sqrt{2E_n \over \hbar \omega},\ {\rm and} \ \ \tilde{z}  \equiv  {z \over z_t}.
    \label{eq: SHO dimensionless variables}
\end{equation}
Here $x_t$ is the turning point, a concept introduced for both the WKB and MAF approximations. This position delineates classically allowed from classically
forbidden regions.
We write the WKB wave function for positive values of $z$ only, since it is even or odd for a symmetric  potential; the final result is
\begin{widetext}
\begin{equation} \label{eq: SHO WKB wavefunction}
\begin{split}
    \psi_{\rm WKB}(z) =  \tilde{D}
    \left\{
    \begin{array}{ll}   \frac{2} {(1 - \tilde{z}^2)^{1/4}} \sin\left[  \frac{\pi}{4} + (n+{1\over 2})({\rm cos}^{-1} (\tilde{z}) - \tilde{z} \sqrt{1-\tilde{z}^2} \right] &  \quad \quad  {\rm for} \ \ 0<z<z_t -\Delta z \quad \quad \quad \ \ \ (R_1)\\
      2 \sqrt{\pi}\left(n + {1 \over 2}\right)^{\frac{1}{6}} \text{Ai}\left[ \sqrt{2}\left(n + {1 \over2}\right)^{\frac{1}{6}} (z-z_t)\right] & \quad \quad  {\rm for} \ \ z_t-\Delta z<z<z_t+\Delta z \ \ \  (R_2)\\
      \frac{1}{(\tilde{z}^2 -1)^{1/4}} \left[\tilde{z} + \sqrt{\tilde{z}^2-1}\right]^{n+\frac{1}{2}}e^{-\frac{z^2}{2} \sqrt{1 - ({z_t \over z})^2}}  & \quad \quad  {\rm for} \ \ z> z_t+\Delta z \quad \quad \quad \quad \ \ \ \ \  (R_3).
\end{array} 
\right.\\
\end{split}
\end{equation}
\end{widetext}
For the SHO, the WKB approximation produces $z_t = \sqrt{2n+1}$, which coincides with the exact result; this has been used in Eq.~(\ref{eq: SHO WKB wavefunction}).
The constant $\tilde{D}$ is found by normalizing the square of the wave function through numerical integration over the entire domain.
Note that a ``patch'' of unknown width $2\Delta z$ has been used to delineate the 3 regions ($R_1$, $R_2$, and $R_3$) for this piecewise defined function. The Airy function ensures
that this wave function, which in its entirety we call the WKB wave function, has no divergence at the turning points. However, the boundaries determined by $\Delta z$ are
ill-defined. As detailed in Appendix \ref{appsec:wkb}, the boundaries arise by a matching process that requires them to be both far away from the turning point
(so the asymptotic expressions used will apply) and close to the turning point (so that linearization of the potential in the vicinity is accurate). 
The actual value of $\Delta z$ does not matter for determining the eigenvalues, but it does matter for determining the eigenfunctions.

Figure~\ref{fig: SHO wavefunctions} shows the WKB wave function (in solid blue) as a function of $z$; only the positive region is shown, as the negative region is a mirror reflection
(in (a) and (c)) or an antisymmetric reflection (in (b)). We have used a definite choice of $\Delta z$, so that the patching region, defined by $R_2$, is shaded in yellow. Note that there
are discontinuities in the wave function, an unacceptable circumstance, that definitely diminishes the ``glow'' of the exact eigenvalue achieved with the same approximation.
Also note that this is unavoidable, as the patching wave function, while hidden under the yellow shaded region, is shown to the left and right of $R_2$ with a curve indicated by
green triangular symbols. This curve is just the Airy function given in the middle line of Eq.~(\ref{eq: SHO WKB wavefunction}).  No choice of $\Delta z$ will make what we call the
bare WKB curve, indicated by blue circular dots and defined by the first and third lines of Eq.~(\ref{eq: SHO WKB wavefunction}), intersect with the curve given by the
green triangles to the left of the turning point. This state of affairs makes the result of exact WKB eigenvalues seem serendipitous, and motivates us to seek an
improved semi-classical approximation, which is the subject of subsection~\ref{subsec: SHO MAF}.

\subsection{The MAF Formalism} \label{subsec: SHO MAF}

In the MAF formalism we use a wavefunction of the form,\cite{MAF_and_WKB_solns,Approx_sol_to_wave_eqn}
\begin{equation} \label{eq: MAF initial wf}
\psi(x) = F(x) Ai(q(x)) + G(x) Bi(q(x)),
\end{equation}
where $Ai(q(x))$ and $Bi(q(x))$ are the two independent solutions of the Airy equation, and are known as Airy functions, and $F(x)$
and $G(x)$ are as-of-yet undetermined functions. Note that $q(x)$ is also an undetermined function.
When we enter Eq.~(\ref{eq: MAF initial wf}) into the Schr\"odinger equation, and make approximations similar to those in the
WKB approximation (see Appendix \ref{sec:appendix_maf}), we obtain
\begin{equation} \label{eq: MAF q(x)}
    q(x) = \pm \left(\frac{3}{2} \int_{x} \sqrt{\frac{2m}{\hbar^2} \left|E-V(x')\right|}dx'\right)^{2/3},
\end{equation}
where $\int_x$ is properly written as $\int_x^{x_t}$ for $x<x_t$ and as $\int_{x_t}^{x}$ for $x>x_t$. The positive or negative
sign is chosen according to the slope of the potential at the turning point $x_t$. As in subsection (\ref{subsec: SHO WKB})
we consider only positive $x$ since the potential is assumed to obey $V(-x) = V(x)$.

Following the procedure outlined in Appendix \ref{sec:appendix_maf}, we obtain

\begin{equation}
    \psi_{\text{MAF}}(x) = C_1 \frac{q(x)^{1/4}}{\sqrt{|k(x)|}} Ai\left(q(x)\right) + C_2 \frac{q(x)^{1/4}}{\sqrt{|k(x)|}} Bi\left(q(x)\right),
    \label{eq: MAF general wavefunction}
\end{equation}
where
\begin{equation}
    k(x) \equiv \sqrt{{2m \over \hbar^2}[E - V(x)]},
\end{equation}
and $C_1$ and $C_2$ are constants.
\begin{figure}[!htb]
     \begin{minipage}[h]{1.0\linewidth}
        \centering
        \includegraphics[width=1.0\linewidth]{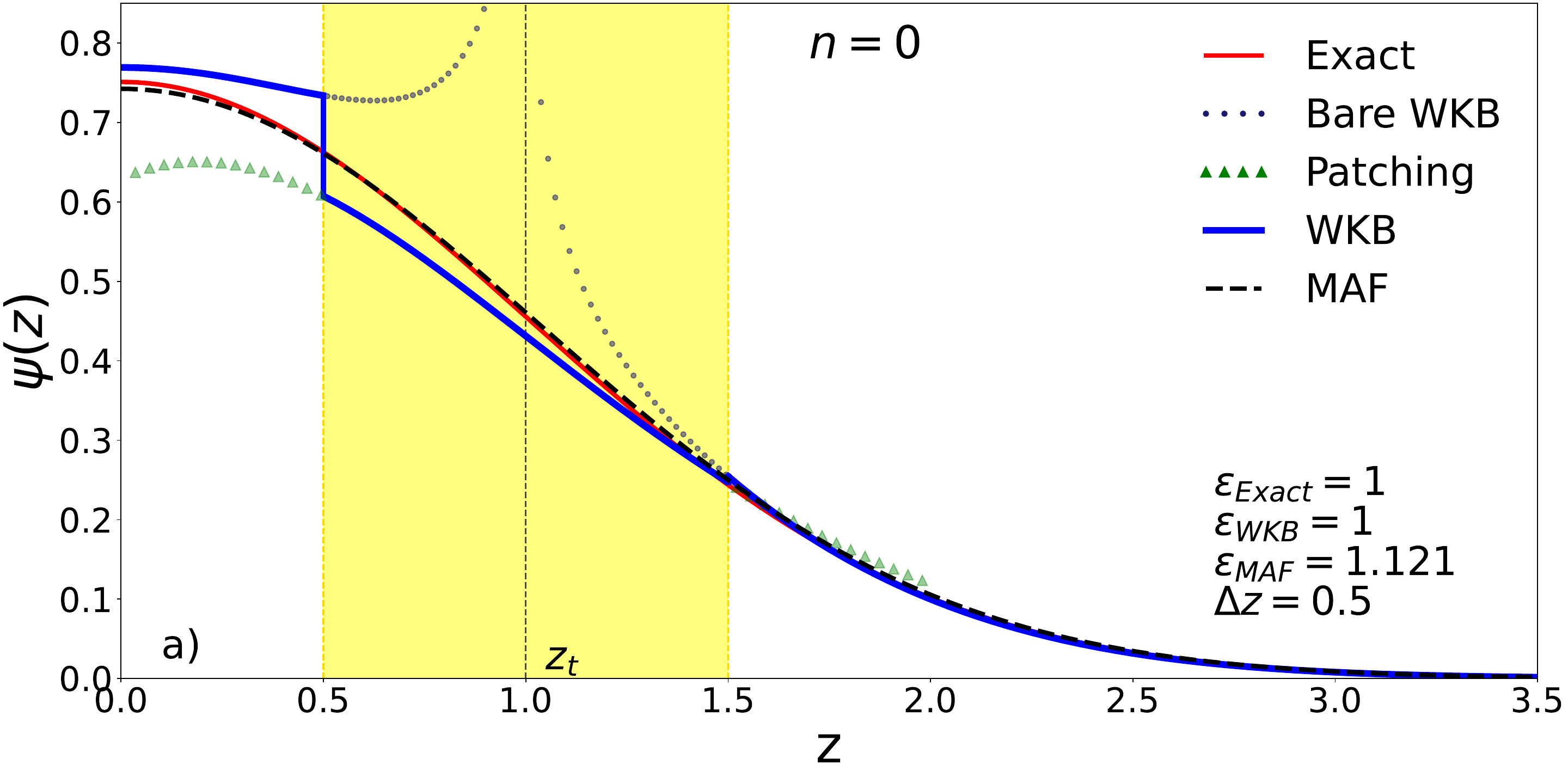}
     \end{minipage}
\vspace{0.00mm}
    \begin{minipage}[h]{1.0\linewidth}
       \centering
       \includegraphics[width=1.0\linewidth]{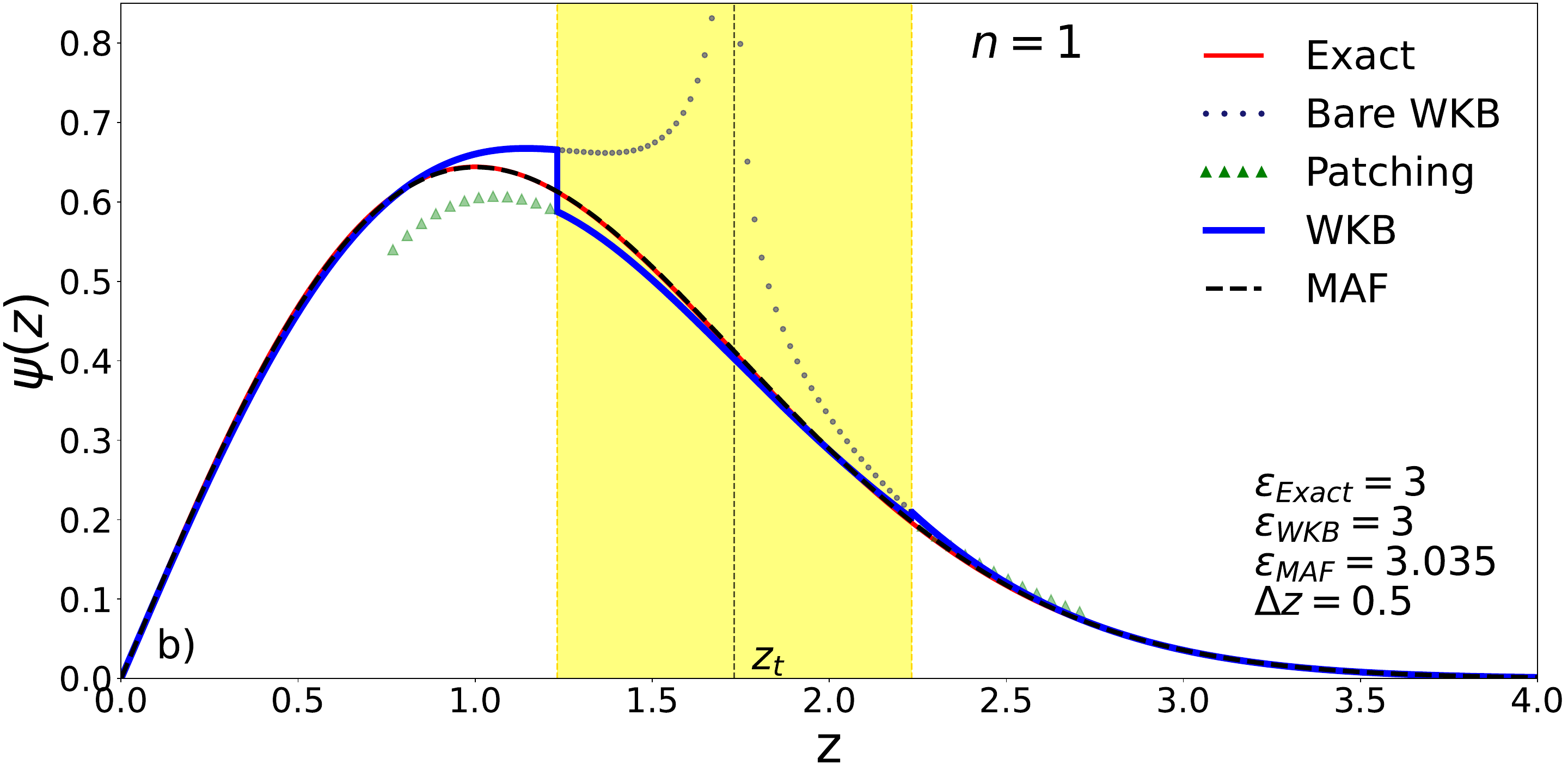}
     \end{minipage}
\vspace{0.00mm}
    \begin{minipage}[h]{1.0\linewidth}
       \centering
       \includegraphics[width=1.0\linewidth]{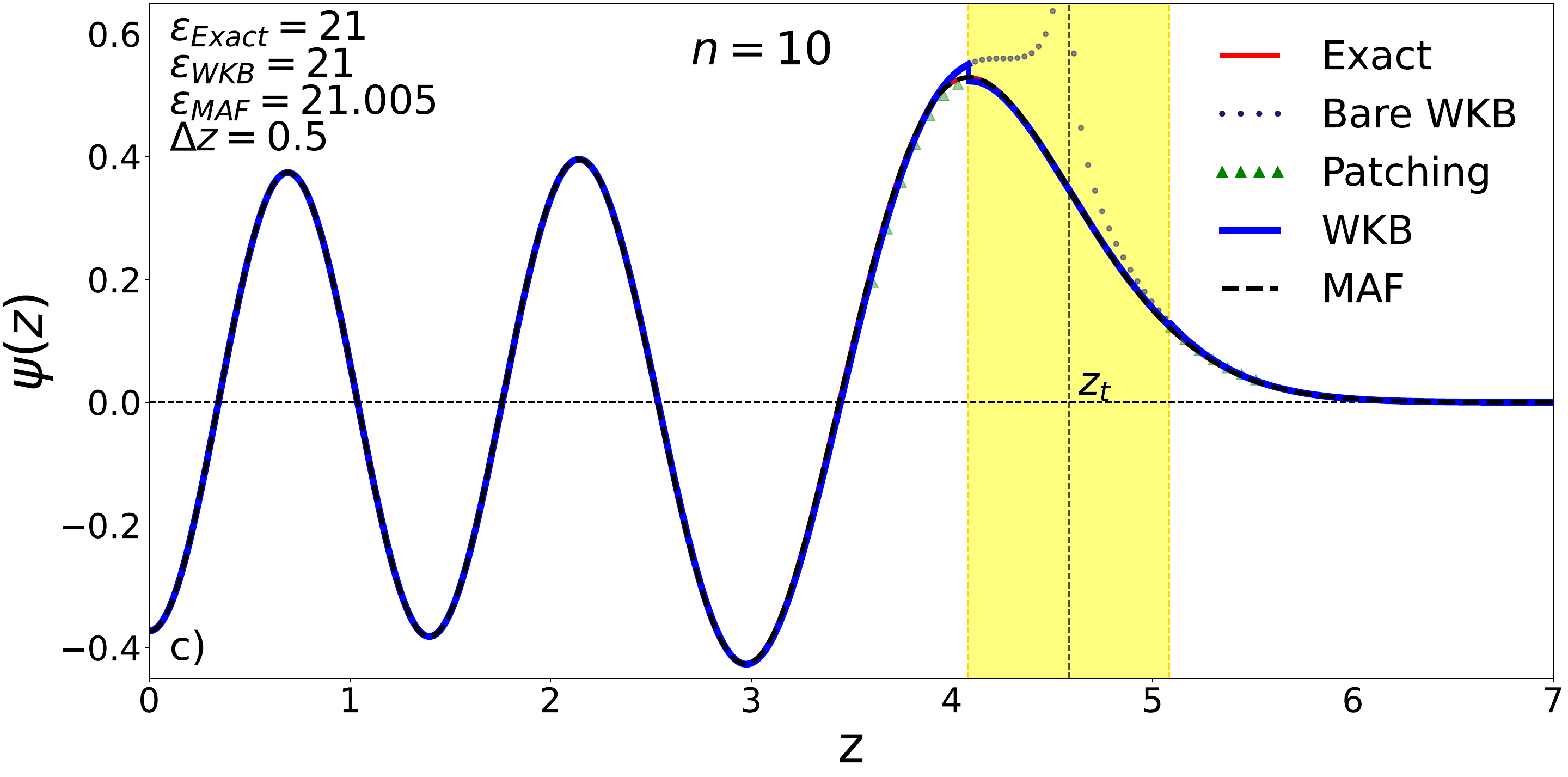}
    \end{minipage}
\caption{Plots of the exact (solid red curve), bare WKB (dotted blue), patching (green triangles), WKB (solid blue curve) and MAF (dashed black curve) wave functions vs. 
$z \equiv x/x_{HO}$ for positive $z$ only, for the (a) ground, (b) first excited, and (c) $10^{\rm th}$ excited state for the SHO potential. 
In each case, yellow-shaded areas indicate patching regions for a specific choice of $\Delta z$, in which the patching wave function is utilized.
These are used to replace the divergent bare WKB functions
in this region, and to interpolate between the bare WKB functions on either side. As explained in the text,
a different choice for $\Delta z$ will not alleviate the discontinuity in the WKB approximation. This plot illustrates that (i) the WKB result is unacceptable, (ii) the 
MAF result is quite accurate, and (iii) the MAF results improve with increasing quantum number.}
\label{fig: SHO wavefunctions}
\end{figure}

Specifically for the SHO, since the wave function must be square-integrable and $Bi(q(x))$ diverges for large $x$, we set $C_{2}=0$.
The SHO wave function can then be written in terms of the dimensionless variables given in Eq. (\ref{eq: SHO dimensionless variables}) for $z>0$:
\begin{equation}
\begin{split}
    \psi_{\text{MAF}}(z) = \tilde{C}&   
    \left\{
\begin{array}{ll}
   \frac{\left(\frac{3}{2}s_1(z) \right)^{\frac{1}{6}}}{ \left(z_t^2-z^2\right)^{\frac{1}{4}}} Ai\left[- \left(\frac{3}{2} s_1(z) \right)^{\frac{2}{3}} \right], &  z<z_t\\
    \frac{\left(\frac{3}{2}s_2(z) \right)^{\frac{1}{6}}}{ \left(z^2-z_t^2\right)^{\frac{1}{4}}} Ai\left[+ \left(\frac{3}{2} s_2(z) \right)^{\frac{2}{3}} \right], &   z>z_t\\
\end{array} 
\right.\\
\end{split}
\label{maf_wavefunctions}
\end{equation}
where ($\tilde{z} \equiv z/z_t$)
\begin{eqnarray}
    s_1(z) &\equiv & \int_z^{z_t} \sqrt{z_t^2-z'^2}dz' \nonumber \\
    &= & \frac{z_t^2}{2} \left[ {\rm cos}^{-1}(\tilde{z}) - \tilde{z} \sqrt{1 - \tilde{z}^2}\right],\\
    s_2(z) &\equiv & \int_{z_t}^{z} \sqrt{z'^2-z_t^2}dz' \nonumber \\
    &= & \frac{z_t^2}{2} \left[ \tilde{z} \sqrt{\tilde{z}^2 - 1} - \ln \left(\tilde{z} + \sqrt{\tilde{z}^2 - 1}\right) \right].
    \label{definitions2}
    \end{eqnarray}
As was the case with the WKB approximation, eigenvalues (i.e. $z_t^2$) for wave functions with even parity are determined by

\begin{equation} \label{eq: general eigenvalue condition even}
    \psi'(0) = 0 \ \ \Rightarrow  \ \ \frac{Ai'(q(0))}{Ai(q(0))} = \frac{1}{2} \frac{q''(0)}{q'^2(0)} , \  \text{(even states)}
\end{equation}
while eigenvalues for wave functions with odd parity are determined by
\begin{equation} \label{eq: general eigenvalue condition odd}
\psi(0) = 0  \ \   \Rightarrow \ \ Ai(q(0)) = 0.     \quad \quad  \quad \quad    \text{(odd states)}
\end{equation}
By solving these equations we find the dimensionless energies,
\begin{equation} \label{eq: SHO MAF energies}
      z_t^2 = 1.121, \ \ 3.035, \ \ 5.023, \ \ 7.016, \ \ 9.013 \dots
\end{equation}
As expected, these slowly approach the odd integers as the quantum number increases. 

In contrast to the WKB wave function for the ground state of the SHO shown in Fig.~\ref{fig: SHO wavefunctions}(a), (i) the MAF wave function does not diverge
at the turning point, (ii) no patching region is required, and (iii) the MAF very accurately matches the exact result over the entire domain.
The MAF wave function improves with increasing quantum number, as shown in (b) and (c), and is already indistinguishable by eye from the exact result for the first
excited state.

\section{The Double-Well Potential} \label{sec:maf}

\begin{figure}[tp]
\begin{center}
\includegraphics[height=2.45in,width=3.3in]{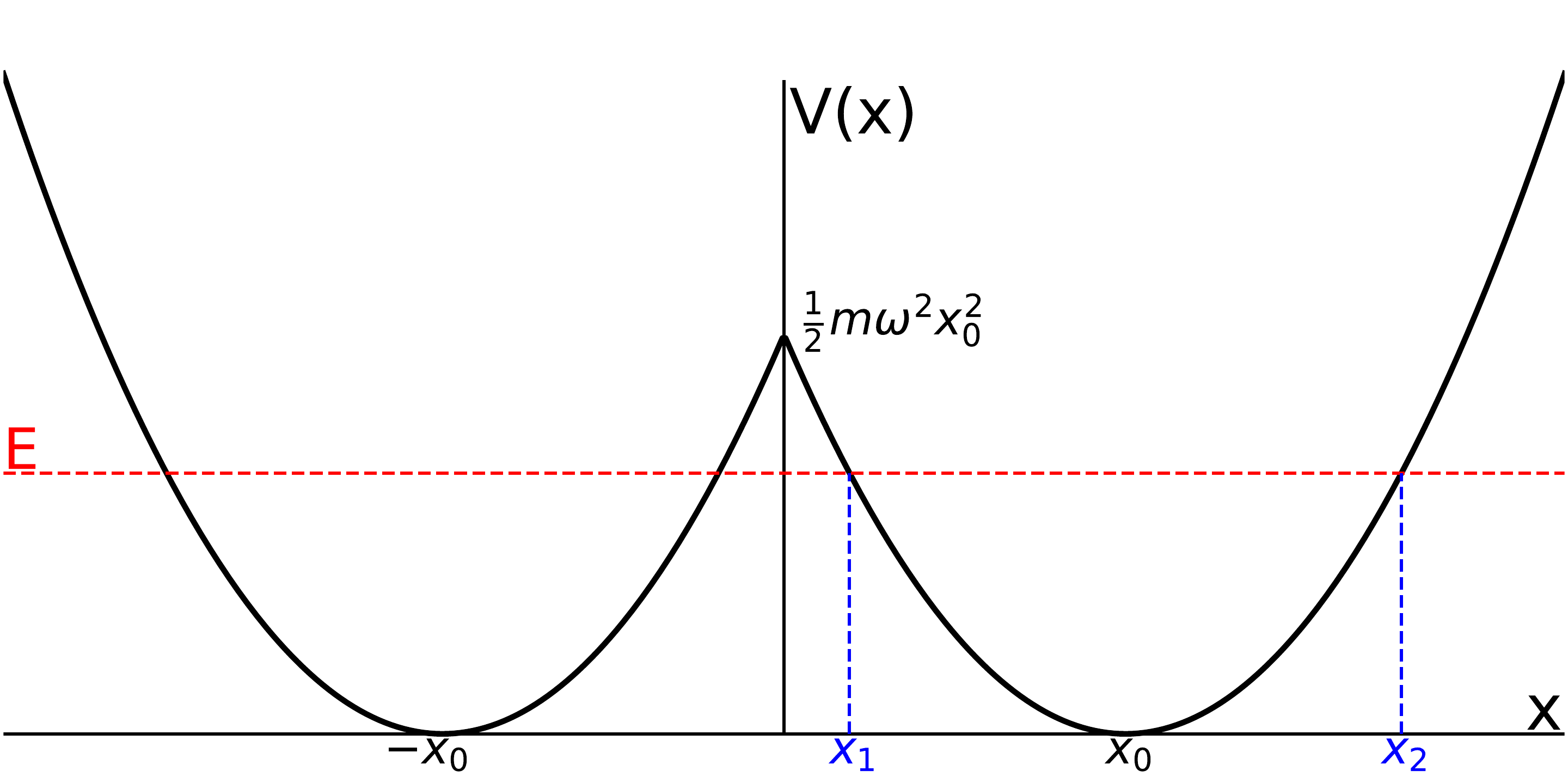}
\end{center}
\caption{The Double-Well Potential described by Eq.~(\ref{vdwp}). Note the value at the cusp, $V(0) = {1 \over 2}m\omega^2 x_0^2$.
We also indicate an energy level denoted by $E$, which defines the two turning points for $x>0$, $x_1$ and $x_2$.}
\label{fig2}
\end{figure}

While the MAF approximation works very well for the SHO, we want to test its accuracy with another important potential in quantum mechanics, the double-well potential.
As discussed Ref.~\onlinecite{Double-potential-well}, there are many forms for this potential that one can adopt. For simplicity, here we will use the form
\begin{equation}
    V(x) = \frac{1}{2} m\omega^2 \left(|x|-x_0\right)^2,
    \label{vdwp}
\end{equation}
which describes two displaced harmonic oscillator potentials with a cusp between them forming a potential barrier. A sketch of the potential is given in Fig.~\ref{fig2}.
As in Section (\ref{sec:sho}), we consider positive $x$ only and work with dimensionless coordinates, $z \equiv x/x_{HO}$, with $x_{HO} \equiv \sqrt{\hbar/(m\omega)}$.
The cusp value at the origin in dimensionless units is given by $v(0) \equiv V(0)/(\hbar \omega/2) = z_0^2$, where $z_0 \equiv x_0/x_{HO}$. There are two distinct
physical regimes for this potential. Here we will consider only bound states with dimensionless energy less than $z_0^2$, as indicated by the horizontal red dashed
line in the figure. Then for the purposes of the approximate treatments, there are a total of four turning points (i.e. two for positive values of $x$). As the value of $z_0^2$
increases, we expect the ground state to approach near-degeneracy, with the usual symmetric and antisymmetric
combinations of the ground state for each well.

There are several additional complexities for this potential compared to the SHO.  First, while exact analytical solutions are possible, they
require confluent hypergeometric functions. We have opted to utilize numerically exact solutions using matrix mechanics, a full description of
which is given in Refs.~\onlinecite{Matrix_mechanics,Double-potential-well}. 
For completeness, this methodology is briefly described in Appendix~\ref{sec:appendix_mat},
and the reader is referred to this or the references just cited. However, an additional complication is the presence of two turning points for $z>0$, requiring
a piecewise function for both the WKB and MAF approximations. In this case, however, both the wave function and its derivative will be continuous. These positive turning
points are given by $z_1 = z_0 - \sqrt{\epsilon}$ and $z_2 = z_0 + \sqrt{\epsilon}$, where $\epsilon \equiv E/(\hbar \omega/2)$ is the dimensionless energy.
Finally, the presence of a cusp in the potential
poses a difficulty in implementing both the WKB and the MAF formulations; however, by requiring the continuity of the derivative of the wave function
at the origin, this difficulty is resolved.

\subsection{The WKB Approximation for the DWP} \label{subsec: DWP WKB}

The presence of two turning points for $z>0$ requires five distinct regions, labeled below from $R_1$ to $R_5$.
With all details relegated to the Appendix, we write down the final result in dimensionless variables, for $z> 0$,
\begin{widetext}
\begin{eqnarray} \label{eq: DPW WKB final wavefunction}
    &\psi_{\text{WKB}}(z) = \tilde{D}
    &\left\{
\begin{array}{llll}
\frac{1}{\left(\left(z-z_0\right)^2-\epsilon\right)^{1/4}} \left(2\cos\left(\frac{\pi \epsilon}{2}\right) e^{w_1(z)} + \sin\left(\frac{\pi \epsilon}{2}\right) e^{-w_1(z)} \right), & \ \ \ \ \ \ \ \ \ 0<z<z_1-\Delta z \ \ \ \ (R_1)\\
    \sqrt{\frac{4\pi}{\alpha}} \left[\sin\left(\frac{\pi \epsilon}{2}\right) Ai(\alpha(z-z_1)) + \cos\left(\frac{\pi \epsilon}{2}\right) Bi(\alpha(z-z_1))\right], & z_1-\Delta z<z<z_1 + \Delta z \ \ \ \ (R_2) \\
    \frac{2}{\left(\epsilon-\left(z-z_0\right)^2\right)^{1/4}} \sin(w_3(z) + \frac{\pi}{4}), & z_1+\Delta z < z < z_2-\Delta z \ \ \ \ (R_3)\\
    \sqrt{\frac{4\pi}{\alpha}} Ai(\alpha (z-z_2)), & z_2-\Delta z < z < z_2 + \Delta z \ \ \ \ (R_4) \\
    \frac{1}{\left(\left(z-z_0\right)^2-\epsilon\right)^{1/4}}e^{-w_4(z)}, & z_2+\Delta z < z\ \ \ \ \ \ \ \ \ \ \ \ \ \ \ \ \ \ (R_5) \\
\end{array} 
\right.
\end{eqnarray}
where the auxiliary functions
\begin{eqnarray}
    &\ \ \ \ \ \ \ \ \ \ \ w_1(z) \equiv  \int_z^{z_1} \sqrt{(z'-z_0)^2-\epsilon}\ dz' =
    \frac{1}{2}\Bigg( \ (z_0-z)\sqrt{(z-z_0)^2-\epsilon} \ +\ \epsilon \ln  \left(\frac{z_0-z-\sqrt{(z-z_0)^2-\epsilon}}{\sqrt{\epsilon}} \right) \Bigg),\nonumber \\
    &w_2(z)  \equiv \int_{z_1}^{z} \sqrt{\epsilon-(z'-z_0)^2}\ dz' \ =
    \frac{1}{2}\bigg( \ (z-z_0) \sqrt{\epsilon-(z-z_0)^2} \ +\ \epsilon \arccos \left( \frac{z_0-z}{\sqrt{\epsilon}} \right) \bigg),\nonumber \\
    &w_3(z)  \equiv  \int_z^{z_2} \sqrt{\epsilon-(z'-z_0)^2}\ dz' =
    \frac{1}{2} \bigg( \ (z_0-z) \sqrt{\epsilon-(z-z_0)^2} \ +\ \epsilon \arccos \left( \frac{z-z_0}{\sqrt{\epsilon}} \right) \bigg),\nonumber \\
    &\ \ \ \ \ \ \ \ \ \  w_4(z) \equiv  \int_{z_2}^{z} \sqrt{(z'-z_0)^2-\epsilon}\ dz' \ =
    \frac{1}{2} \Bigg(\  (z-z_0)\sqrt{(z-z_0)^2-\epsilon} \ -\
    \epsilon \ln  \left( \frac{z-z_0+\sqrt{(z-z_0)^2-\epsilon}}{\sqrt{\epsilon}} \right) \Bigg)
    \label{w_definitions}
    \end{eqnarray}
are used here and below. Note that for simplicity we have used the same half-width, $\Delta z$, for both patching regions.
As done previously, odd or even solution eigenvalues are obtained by demanding that the wave function or
its derivative be zero, respectively, at $z=0$. Therefore only the first line of Eq.~(\ref{eq: DPW WKB final wavefunction}) is required. By enforcing
Eqs.~(\ref{eq: general eigenvalue condition odd}) and (\ref{eq: general eigenvalue condition even}), respectively, we obtain
the equations
\begin{equation} \label{eq: DWP WKB eigenvalue equations}
\begin{split}
&\cot\left(\frac{\pi \epsilon}{2}\right) = \frac{1}{2} g_{\rm odd}(z_0,\epsilon)  \left( \frac{z_0 +\sqrt{z_0^2 -\epsilon}}{\sqrt{\epsilon}} \right)^\epsilon e^{-z_0\sqrt{z_0^2-\epsilon}}\ \ \ \ \ \ \ \ \text{(odd)}\\
&\cot\left(\frac{\pi \epsilon}{2}\right) = \frac{1}{2} g_{\rm even}(z_0,\epsilon)  \left( \frac{z_0 +\sqrt{z_0^2 -\epsilon}}{\sqrt{\epsilon}} \right)^\epsilon e^{-z_0\sqrt{z_0^2-\epsilon}}\ \ \ \ \ \ \ \ \text{(even)}
\end{split}
\end{equation}
\end{widetext}
where
\begin{eqnarray}
g_{\rm odd} \equiv &-1 \nonumber \\
g_{\rm even} \equiv &\left(\frac{2(z_0^2-\epsilon)^{3/2} + z_0}{2(z_0^2-\epsilon)^{3/2} - z_0}\right).
\label{g_definitions}
\end{eqnarray}
Equation~(\ref{eq: DWP WKB eigenvalue equations}) is used to determine the dimensionless eigenvalue, $\epsilon$, and then this is inserted into
Eq.~(\ref{eq: DPW WKB final wavefunction}) to determine the wave function. As before, the constant $\tilde{D}$ is determined by normalizing the wave function.
Note that we have accounted for the non-zero negative slope of the potential at $z = 0^+$, and this results in the form of $g_{\rm even}$ as given. One might be tempted to
set this slope to zero, with the argument that the {\it average} slope at the cusp of the potential is zero, in which case $g_{\rm even}$ reduces to unity. This argument
is incorrect, however; when we tried to use a zero slope for the potential, the wave function had a cusp at the origin (discontinuity in the derivative of the wave function), which is both inconsistent and unacceptable.

Results for the WKB approximation will be shown below, after we determine the results for this potential in the MAF approximation.\\

\subsection{The MAF Formalism for the DWP} \label{subsec: DWP MAF}

As mentioned in the previous section, the double-well potential results in an increased number of turning points. This means
that an MAF approximation is required for each turning point. These approximations must be matched to one another at some
intermediate point. The natural point for doing this is at the well minimum, where $z=z_0$. As outlined in Appendix \ref{sec:appendix_maf}, requiring
continuity of the wave function and its derivative will fix two of the coefficients. Requiring square integrability as $z \rightarrow \pm \infty$ will eliminate another,
which leaves one overall coefficient to be determined by normalizability. The energies can be determined by the first equality of
Eqs.~(\ref{eq: general eigenvalue condition even}) or (\ref{eq: general eigenvalue condition odd}) for eigenstates with even or odd parity, respectively.
Then the wave function can be written as (see Appendix \ref{sec:appendix_maf})
\begin{widetext}
\begin{eqnarray} \label{eq: DWP MAF final wavefunction}
    \psi_{\text{MAF}}(z) =  \tilde{C}
    \left\{
\begin{array}{ll}
    \frac{ \left(\frac{3}{2} w_1(z)\right)^{\frac{1}{6}}}{\left(\left(z-z_0\right)^2-\epsilon\right)^{\frac{1}{4}}} \left(c_1 Ai \left((+\frac{3}{2}w_1(z)^{\frac{2}{3}} \right) + c_2 Bi \left((+\frac{3}{2}w_1(z)^{\frac{2}{3}} \right)\right),
    & \ \ \ \ \ \ \ \ \ 0<z<z_0 \\
    \frac{\left(\frac{3}{2} w_2(z)\right)^{\frac{1}{6}}}{\left(\epsilon-\left(z-z_0\right)^2\right)^{\frac{1}{4}}} \left(c_1 Ai \left(-(\frac{3}{2}w_2(z)^{\frac{2}{3}} \right) + c_2 Bi \left(-(\frac{3}{2}w_2(z)^{\frac{2}{3}} \right)\right),
    & \ \ \ \ \ \ \ \ \ z_1 <z<z_0  \\
    \frac{ \left(\frac{3}{2} w_3(z)\right)^ {\frac{1}{6}}}{\left(\epsilon-\left(z-z_0\right)^2\right)^{\frac{1}{4}}} Ai \left(-(\frac{3}{2}w_3(z)^{\frac{2}{3}} \right), &\ \ \ \ \ \ \ \ \ z_0 <z<z_2  \\
    \frac{ \left(\frac{3}{2} w_4(z)\right)^ {\frac{1}{6}}}{\left(\left(z-z_0\right)^2-\epsilon\right)^{\frac{1}{4}}} Ai \left((+\frac{3}{2}w_4(z)^{\frac{2}{3}} \right), &\ \ \ \ \ \ \ \ \ z_2<z \\
\end{array}
\right.
\end{eqnarray}
\end{widetext}
where
\begin{eqnarray} \label{c_coeff}
c_1 &=& 1 + 2 \pi Bi(\gamma) \left( {1 \over 4 \gamma} Ai(\gamma) + Ai^\prime(\gamma)\right) \nonumber \\
c_2 &=& \ \ \ -2 \pi Ai(\gamma) \left( {1 \over 4 \gamma} Ai(\gamma) + Ai^\prime(\gamma)\right) 
\end{eqnarray}
with
\begin{equation}
\gamma = -{1 \over 4} (3 \pi \epsilon)^{2/3}.
\label{gamma}
\end{equation}
\begin{figure}[!htb]
\centerline{\includegraphics[scale=0.17]{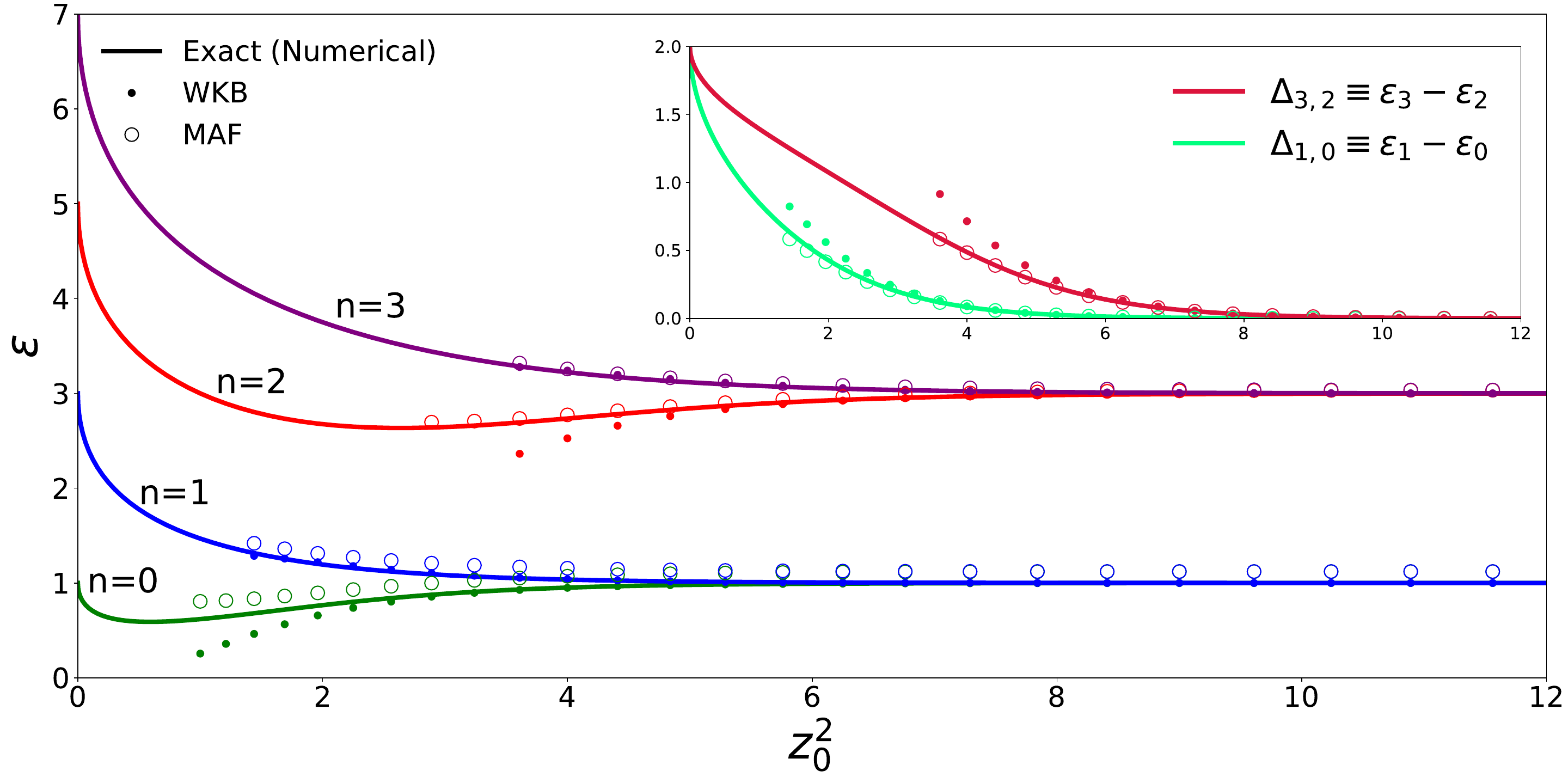}}
\caption{Plot of the first four energy levels of the DWP showing exact, WKB and MAF dimensionless energies ($\epsilon$) vs $z_0^2 \equiv V(0)/(\hbar \omega/2)$.
For small values of $z_0^2$ no eigenvalues exist below the central maximum, and so we do not show WKB and MAF results for these cases.
For a large enough central barrier height $z_0^2$, the DWP approaches two independent harmonic oscillators, so the even/odd parity pairs of states 
approach near-degeneracy (see Tables~\ref{tab: DWP WKB Numeric eigvals} and \ref{tab: DWP MAF Numeric eigvals} in Appendix~\ref{sec:appendix_tables} ). 
The inset shows the energy difference ($\Delta_{3,2}$ and $\Delta_{1,0}$) between the even and odd parity state energies, and the MAF approximation accurately
reproduces the exact result.
}
\label{fig: DWP first 4 energies}
\end{figure}

\begin{figure}[!htb]
     \begin{minipage}[h]{0.9\linewidth}
        \centering
        \includegraphics[width=0.9\linewidth]{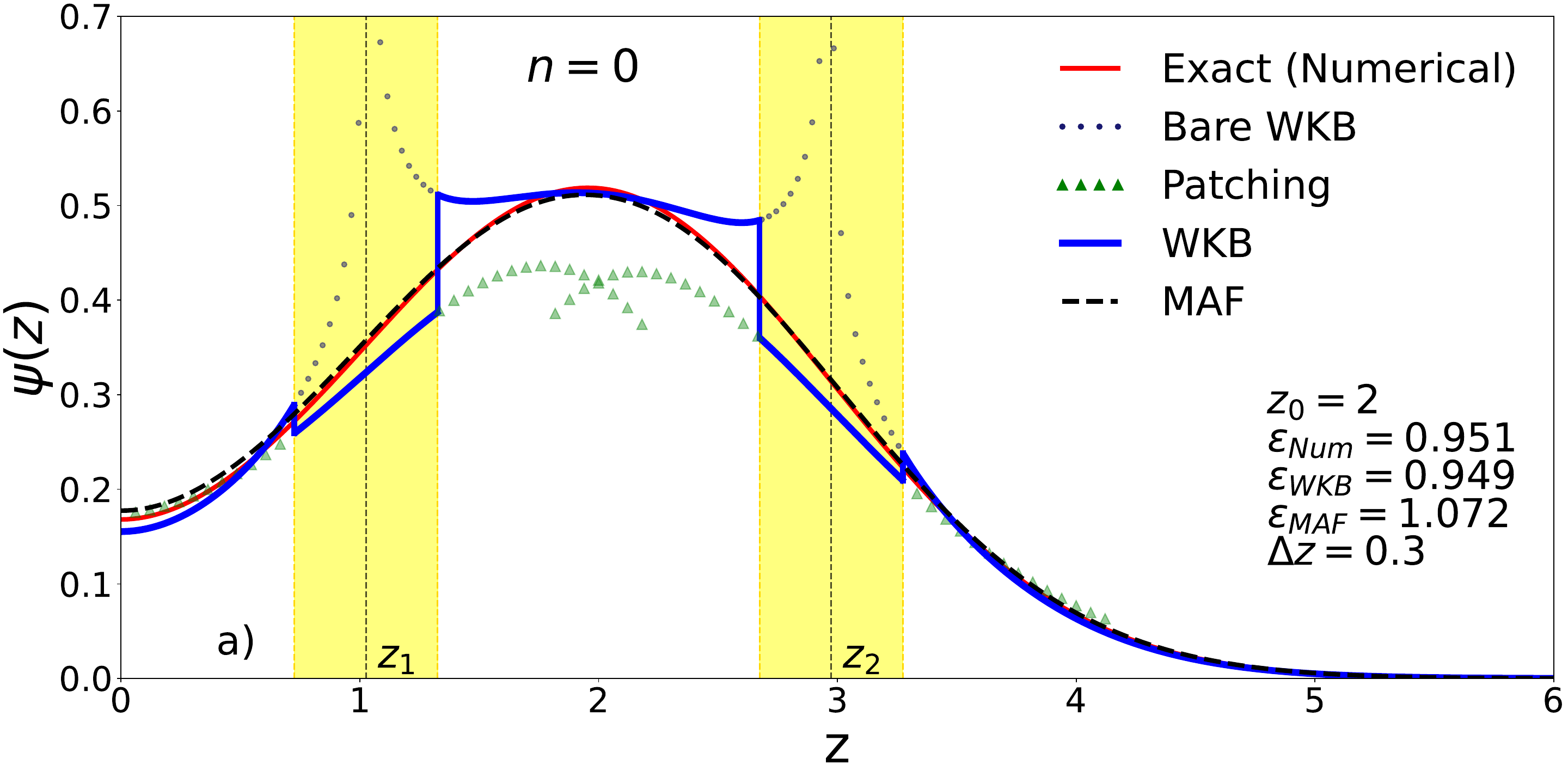}
     \end{minipage}
\vspace{0.00mm}
    \begin{minipage}[h]{0.9\linewidth}
       \centering
       \includegraphics[width=0.9\linewidth]{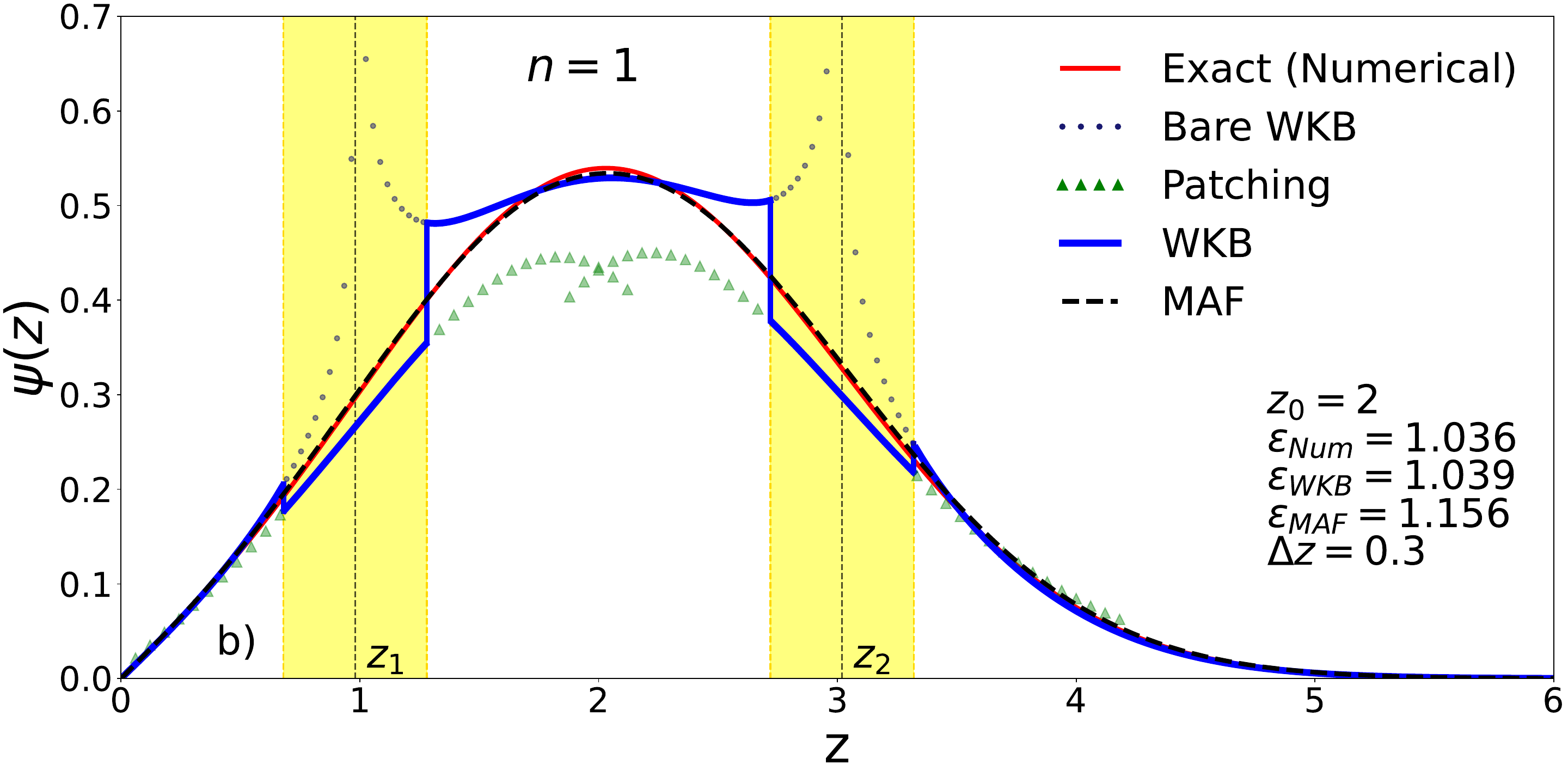}
     \end{minipage}
\vspace{0.00mm}
    \begin{minipage}[h]{0.9\linewidth}
       \centering
       \includegraphics[width=0.9\linewidth]{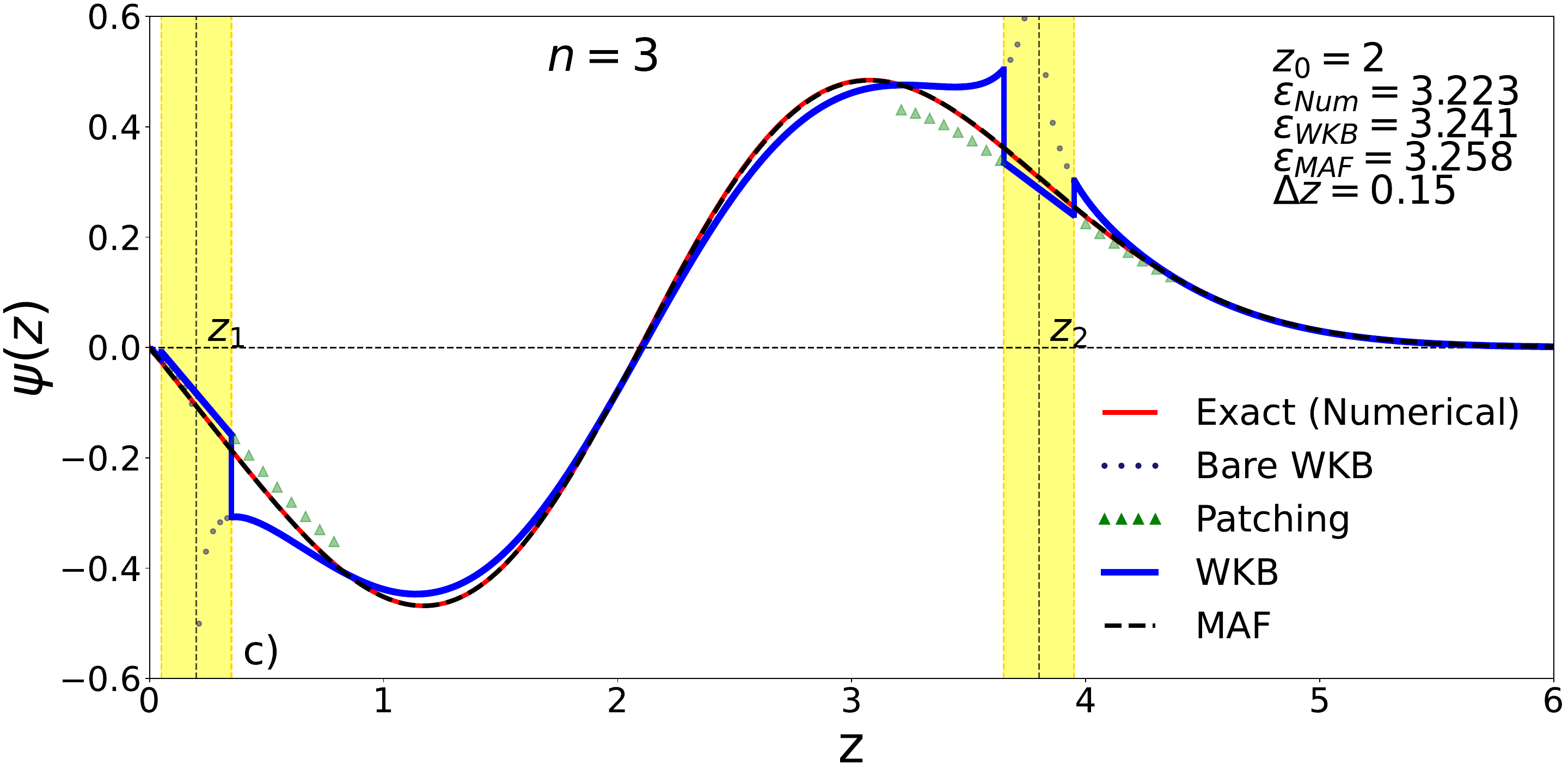}
    \end{minipage}
\caption{Plots of the exact (solid red curve), bare WKB (dotted blue), patching (green triangles), WKB (solid blue curve) and MAF (dashed black curve) wave functions vs. 
$z \equiv x/x_{HO}$ for positive $z$ only, for the (a) ground, (b) first excited, and (c) third excited state for the DWP potential. As before, patching regions are
indicated by shaded yellow areas in each case, for a specific choice of $\Delta z$. As explained in the text,
a different choice for $\Delta z$ will not alleviate the discontinuity in the WKB approximation. As was the case with the SHO, this plot illustrates that
(i) the WKB result is unacceptable, (ii) the MAF result
is quite accurate, and (iii) the MAF results improve with increasing quantum number.}
\label{fig: DWP wavefunctions}
\end{figure}
These expressions for $c_1$ and $c_2$ are determined by requiring
continuity of the wave function and its derivative at the point $z_0$. Substituting $z=z_0$ into the second and third lines of Eq.~(\ref{eq: DWP MAF final wavefunction}),
and requiring that they equal one another gives one equation, while equating their derivatives gives a second equation. The solution of these two equations yields
Eq.~(\ref{c_coeff}). Note that we used the fact that the Wronskian of the two Airy functions is a constant, $1/\pi$.

Results for the eigenvalues for the various approximations are shown in Fig.~\ref{fig: DWP first 4 energies}. One can see that the
MAF approximation improves over the WKB approximation for smaller values of $z_0^2$. At larger values of $z_0^2$, the double-well potential becomes
essentially decoupled single oscillators, so the WKB eigenvalues are more accurate.

Finally, we show the exact (solid red), WKB (solid blue) and MAF (dashed black) wave functions as a function of $z \equiv x/x_{HO}>0$ in Fig.~\ref{fig: DWP wavefunctions}
for (a) $n=0$, (b) $n=1$ and (c) $n=3$. We used an intermediate value of $z_0^2 = 4$ so that the the first two wave functions are reasonably distinct (aside from $n=0$ ($n=1$)
being symmetric (antisymmetric)). Also shown are the patching portions (green triangles) and the bare WKB (blue dots). The latter shows the characteristic divergence 
at the turning points. As was the case in the SHO, here the discontinuities in the WKB wave function are unacceptable, and the MAF wave function is very accurate.

\section{Summary}

When the WKB approximation is presented in textbooks, \cite{Griffiths_3rd_edition} a great deal of effort is given to describe the patching of the wave function in order
to justify the connection formula that connects the wave functions on either side of the turning point. This allows one to determine the energy eigenvalues, and assess
these with respect to the exact results for simple models like the SHO. But one almost never examines the actual wave function that results from this procedure. Implicit
in the procedure that uses asymptotic expressions for the functions involved, however, is that these components of the wave function, i.e. the bare WKB and the patching
Airy functions, {\it do} connect. As we have illustrated in this paper, these functions {\it do not} connect, and the resulting WKB wave function inevitably has
discontinuities.

A remedy to this unacceptable situation is the Modified Airy Function approximation.\cite{MAF_and_WKB_solns,Approx_sol_to_wave_eqn} Because the MAF
approximation is not well known, we have first applied the methodology to the Simple Harmonic Oscillator potential, both to illustrate the deficiencies in the WKB, and
to demonstrate how well the MAF works. While the MAF eigenvalues for the SHO are no longer exact, the MAF wave functions are remarkably accurate, even for the
ground state.

We followed the same path for a more interesting potential, the double-well potential, particularly in the form used in Ref.~\onlinecite{Double-potential-well}, consisting
of two displaced harmonic oscillator potentials. Ref.~\onlinecite{Double-potential-well} highlighted the accuracy of the WKB eigenvalues, but did not critically
examine the WKB wave functions. We have shown here that the WKB wave functions have the same unacceptable deficiency as was the case with the SHO --- discontinuities
are unavoidably present in the WKB wave function. We illustrated that the MAF approximation not only remedies this feature of the wave functions, 
but also provides a very accurate wave function when compared to exact numerical solutions.

We have previously used the MAF approximation in a senior undergraduate course at our university. While the technical aspects are challenging, the present generation of
students is capable of accessing a variety of computational resources for which the Airy function, for example, can be accessed as readily as a sine function can be used from a 
scientific calculator. Students are required to understand some aspects of the Airy function anyways when learning the WKB approximation (e.g. their asymptotic limits);
the MAF requires an extra step of finding the zeros of these functions and their derivatives. Moreover, they adopt a much more hands-on approach to dealing concretely
with these functions, and discover, for example, that while asymptotic expressions can agree with one another in principle, in practice the functions involved cannot actually
be connected as we have discovered here.\\

\begin{acknowledgments}
This work was supported in part by the Natural Sciences and Engineering Research Council of Canada (NSERC) and by an MIF from the Province of Alberta.
\end{acknowledgments}

The authors have no conflicts to disclose.

\vskip0.2in

\appendix \label{sec:appendix}

\section{WKB Approximation} \label{appsec:wkb}

The Schr\"odinger equation can be written as
\begin{equation}
{d^2\psi(x) \over dx^2} = -k^2(x) \psi(x),
\label{schro}
\end{equation}
where
\begin{equation}
    k(x) = \sqrt{{2m \over \hbar}[E-V(x)]}.
    \label{kx}
\end{equation}
Turning points are defined by $E=V(x_t)$; when $E~-~V(x)$~is greater than zero, then the particle is classically allowed, whereas when this quantity is less
than zero then the particle is classically forbidden.
The WKB approximation\cite{Griffiths_3rd_edition} begins by separating the wave function into two real functions, an amplitude $A(x)$, and a phase, $\phi(x)$,
\begin{equation}
\psi(x) = A(x) e^{i\phi(x)}.
\label{wkb1}
\end{equation}
Upon making the one approximation, $A^{\prime \prime}(x) \approx 0$, we can solve for $A(x)$ and $\phi(x)$ using Eq.~(\ref{schro}), and we obtain a piecewise WKB wave function,
\begin{widetext}
\begin{eqnarray}
\psi_{\text{WKB}}(x) &\approx &\frac{A}{\sqrt{k(x)}} e^{i \int_x^{x_t} k(x^\prime) dx^\prime} \ +\ \frac{B}{\sqrt{k(x)}} e^{-i \int_x^{x_t} k(x^\prime) dx^\prime}, \ \ \ \ \ \ \ \ \ E > V(x), \ \ \ \ \text{($x< x_t$, allowed)} \label{w1}\\
\psi_{\text{WKB}}(x) &\approx &\frac{C}{\sqrt{|k(x)|}} e^{ \int_{x_t}^{x} |k(x^\prime)| dx^\prime} + \frac{D}{\sqrt{|k(x)|}} e^{- \int_{x_t}^{x} |k(x^\prime)| dx^\prime}. \ \ \ \ \ \ \ \ E < V(x), \ \ \ \ \text{($x> x_t$, forbidden)} \label{w2}
\end{eqnarray}
\end{widetext}
where $A$, $B$, $C$, and $D$ are constants. Given that the wave function has to converge as $x\rightarrow \infty$, then we set $C=0$.
These two wave functions clearly diverge at the turning points. Thus we need a ``patching region'', comprising $x_t \pm \Delta x$, where $2\Delta x$ is some unknown width.
 In this region, assumed to be small enough that a linear approximation to the potential,
\begin{equation}
    V(x) \approx E + V'(x_t) (x-x_t),
\end{equation}
is accurate, the TISE becomes the Airy equation, with Airy function solutions, although we discard $Bi$ since it diverges as $x\rightarrow \infty$.
Then, adopting an asymptotic expansion for the remaining Airy function, we can match coefficients with the bare WKB result from Eq.~(\ref{w2}) and in turn
with Eq.~(\ref{w1}) in order to connect the two solutions. The final result, after performing the integrals required, is given in Eq.~(\ref{eq: SHO WKB wavefunction}).

The eigenvalues are determined by demanding that the wave function at the origin be zero (for wave functions that have odd parity), or that the derivative
of the wave function at the origin be zero (for wave functions that have even parity). Equation~(\ref{eq: SHO WKB wavefunction}) already has the eigenvalue solution
inserted; at an earlier stage the first line reads (for $0<z<z_t -\Delta z$),
\begin{equation}
\psi_{\rm WKB}(z) \frac{2} {(1 - \tilde{z}^2)^{1/4}} \sin\left[  \frac{\pi}{4} + {z_t^2 \over 2}\left({\rm cos}^{-1} (\tilde{z}) - \tilde{z} \sqrt{1-\tilde{z}^2}\right) \right],
 \label{app_earlier}
 \end{equation}
where $z\equiv x/x_{\rm HO}$, $z\equiv x/x_{\rm HO}$, and $\tilde{z} \equiv z/z_t$. This will be zero if the argument of the sine function is an integer multiple of $\pi$
at $z=0$, that is,
\begin{equation}
{\pi \over 4} + {z_t^2 \over 2} {\pi \over 2} = n\pi, \ \ \ \ \text{where} \ \  n = 1,2,3, ....
\label{app_odd}
\end{equation}
This requires $z_t^2 = 4 n - 1$, or $E = {\hbar \omega \over 2}(4n -1), \ n = 1,2,3,...$, which corresponds to
the exact odd parity solutions for the SHO.

For the even parity solutions one must set $\psi_{\rm WKB}^\prime(0) = 0$, which requires the argument of the resulting
cosine function to be a half-integer multiple of $\pi$
at $z=0$, that is
\begin{equation}
{\pi \over 4} + {z_t^2 \over 2} {\pi \over 2} = (2n-1){\pi \over 2}, \ \ \ \ \text{where} \ \  n = 1,2,3, ....
\label{app_even}
\end{equation}
This requires $z_t^2 = 4 n - 3$, or $E = {\hbar \omega \over 2}(4n -3), \ n = 1,2,3,...$, which corresponds to
the exact even parity solutions for the SHO.

For the DWP the algebra is more complicated, as there are two turning points for $x>0$. Then the WKB approximation has to be applied separately for both turning points.
For the turning point on the far right ($x_2$) the argument for the wave function just given applies, whereas for the turning point on the left ($x_1$) a new patching
wave function is required. If we start from scratch, the result is
\begin{widetext}
\begin{equation} \label{eq: DPW WKB general wavefunction}
\begin{split}
    &\psi_{\text{WKB}}(x) = \left\{
\begin{array}{ll}
    \frac{1}{\sqrt{|k(x)|}} \left(A_1 e^{\int_x^{x_1-\Delta x} |k(x')|dx'} + B_1 e^{-\int_x^{x_1 - \Delta x} |k(x')|dx'} \right)  &\ \ \ \ \ \ \ \ \ \ 0<x<x_1-\Delta x \quad \quad \quad (R_1) \\
    \\
    a_2 Ai[-\alpha (x-x_1)] + b_2 Bi[-\alpha (x-x_1)]  &x_1-\Delta x<x<x_1 + \Delta x \ \ \ \ \ \ \ \ \ \ (R_2)\\
    \\
     \frac{1}{\sqrt{k(x)}} \left(A_3 e^{+i\int_{x_1+\Delta x}^x k(x')dx'} + B_3 e^{-i\int_{x_1 + \Delta x}^x k(x')dx'} \right) & x_1+\Delta x <x<x_2 - \Delta x \quad \quad \quad \ (R_3) \\
     \\
    a_4 Ai[\alpha (x-x_2)] + b_4 Bi[\alpha (x-x_2)] & x_2-\Delta x < x < x_2 + \Delta x \ \ \ \ \ \ \ \ \ \ (R_4)\\
    \\
        \frac{1}{\sqrt{|k(x)|}} \left(A_5 e^{\int_{x_2+\Delta x}^x |k(x')|dx'} + B_5 e^{-\int_{x_2 + \Delta x}^x |k(x')|dx'} \right),  & x_2+\Delta x < x\quad \quad \quad \ \ \ \ \ \ \ \ \ \ \ \ \ \ \ \ (R_5) \\
\end{array} 
\right.\\
\end{split}
\end{equation}
\end{widetext}
where we have introduced the patching regions, both with width $2\Delta x$, centred around the two turning points, $x_1$ and $x_2$.
For clarity we have written the result in terms of coordinates with dimension of length, $x$, etc. and used region labels ($R_j, j = 1, ..5$) that will
apply when dimensionless coordinates ($z = x/x_{\rm HO}$, etc.) are used. The constant $\alpha$ is related to the slope of the potential at $x_2$
(and in the same way to the negative of the slope at $x_1$), $\alpha = (4\epsilon)^{1/6}/x_{\rm HO}$.

The exercise of connecting all the wave functions in Eq.~(\ref{eq: DPW WKB general wavefunction}) is lengthy but standard (see, for 
example, Ref.~\onlinecite{Griffiths_3rd_edition}, where the procedure for the double-well potential is given in Prob. 9.17). In words, the
first step is to note that the coefficient $A_5$ must be zero to ensure that the wave function is normalizable. Next, the asymptotic form of the
Airy function in the 4th line is to be matched with the asymptotic form of the 5th line in Eq.~(\ref{eq: DPW WKB general wavefunction}). This process
will determine that $b_4 = 0$ and allows one to write $A_3$ and $B_3$ in terms of $B_5$. Continuing in this way, $a_2$ and $b_2$ in the second
line can be connected to the constant associated with the last three lines of Eq.~(\ref{eq: DPW WKB general wavefunction}). Finally, $A_1$ and $B_1$
can be then written in terms of the same constant. The final result, written in terms of dimensionless variables, is given in Eq.~(\ref{eq: SHO WKB wavefunction})
for the SHO and in Eq.~(\ref{eq: DPW WKB final wavefunction}) for the DWP.

\section{The Modified Airy Function Approximation} \label{sec:appendix_maf}

In the Modified Airy Function approximation,\cite{MAF_and_WKB_solns,Approx_sol_to_wave_eqn} we start with Eq.~(\ref{eq: MAF initial wf}), 
which is in principle exact. The SHO potential represents a more straightforward case, so we will describe here the more complicated case with the
Double-Well Potential. Since there are two turning points for $x>0$ (see Fig.~\ref{fig2}), then we require
\begin{equation}
\begin{split}
    \psi_{\text{MAF}}(x) = &   
    \left\{
\begin{array}{ll}
F_1(x) Ai[q_1(x)] + G_1(x) Bi[q_1(x)], & x<x_0\\
F_2(x) Ai[q_2(x)] + G_2(x) Bi[q_2(x)], & x > x_0\\
\end{array} 
\right.\\
\end{split}
\label{app_maf_wavefunctions}
\end{equation}
where arbitrary coefficients are contained within the auxiliary functions, $F_j(x)$ and $G_j(x)$. Since $Bi[q_2(x)]$ will be a growing exponential
function, we can immediately set 
$G_2(x) = 0$.

We will proceed with one of the terms [say, $F_1(x) Ai(q_1(x))$], as the other two terms follow similar algebra, and they are treated independently.
Substitution of this term into the TISE results in
\begin{equation}
F_1^{\prime \prime} Ai + 2F_1^\prime Ai^\prime q_1^\prime + F_1 Ai^\prime q_1^{\prime \prime} + Ai F_1 \left[ q_1(q_1^\prime)^2 + k^2\right],
\label{app_mfa2}
\end{equation}
where we have suppressed  arguments and we have used the Airy equation to eliminate $Ai^{\prime \prime} = q_1Ai$. We now use the freedom
of our functions and choose $q_1$ to eliminate the last term in square brackets,
\begin{equation}
q_1(x) \left({dq_1(x) \over dx}\right)^2 = -k^2(x).
\label{app_mfa3}
\end{equation}
Note that $q_1(x)$ has a sign that is opposite of that of $k^2(x) \equiv {2m \over \hbar^2}[E-V(x)]$. That is, if we are in the 
classically allowed region, $E>V(x)$, then $q(x) < 0$, and vice-versa. So solving Eq.~(\ref{app_mfa3}) for $q_1(x)$ (see Fig.~\ref{fig2} for the location
of the turning point $x_1$),
we have
\begin{eqnarray}
q_1(x) &=& +\left({3 \over 2} \int_x^{x_1} dx^\prime \sqrt{{2m \over \hbar^2}[E-V(x^\prime)]}\right)^{2/3}  x < x_1\ \ \ \ \\
q_1(x) &=& -\left({3 \over 2} \int_{x_1}^{x} dx^\prime \sqrt{{2m \over \hbar^2}[E-V(x^\prime)]}\right)^{2/3} \ \ x > x_1, \ \ \ \ \ \ \ 
\label{app_mfa4}
\end{eqnarray}
whereas for the second turning point, $x_2$ (see Fig.~\ref{fig2} for the location
of the turning point $x_2$), the signs would be different,
\begin{eqnarray}
q_2(x) &=& -\left({3 \over 2} \int_x^{x_2} dx^\prime \sqrt{{2m \over \hbar^2}[E-V(x^\prime)]}\right)^{2/3}  x < x_2\ \ \ \ \\
q_2(x) &=& +\left({3 \over 2} \int_{x_2}^{x} dx^\prime \sqrt{{2m \over \hbar^2}[E-V(x^\prime)]}\right)^{2/3} \ \ x > x_2. \ \ \ \ \ \ \ 
\label{app_mfa5}
\end{eqnarray}
Another way of prescribing the sign choice for $q(x)$ is that it is dependent on the slope of the potential at the turning point (negative at $x_1$,
and positive at $x_2$).
Evaluation of these integrals is straightforward, and the result is provided by Eq.~(\ref{w_definitions}) in the text.

Returning to Eq.~(\ref{app_mfa2}), once $q(x)$ is defined, and therefore the term in braces is eliminated, we make the semi-classical approximation,
$F_1^{\prime \prime} \approx 0$, which then allows the solution for $F_1$ (or $F_2$ or $G_1$). The result is
\begin{equation}
F_1 \sim \left({q_1(x) \over -k^2(x)}\right)^{1/4}, \ F_2 \sim \left({-q_2(x) \over k^2(x)}\right)^{1/4} \ G_1 \sim \left({-q_1(x) \over k^2(x)}\right)^{1/4}.
\label{app_mfa6}
\end{equation}
Allowing for arbitrary constants, substituting Eq.~(\ref{w_definitions}) for the $q_i(x)$, and using dimensionless coordinates results in 
Eq.~(\ref{eq: DWP MAF final wavefunction}) in the main text. To determine the eigenvalues, for the odd-parity eigenstates we use $\psi_{\rm MAF}(z=0) = 0$, i.e.
we find the values of $\epsilon$ that solve
\begin{equation}
c_1Ai\left[{3 \over 2}{w_1(0)}^{2/3}\right] + c_2Bi\left[{3 \over 2}{w_1(0)}^{2/3} = 0\right],  \ \ (odd)
\label{app_mfa7}
\end{equation}
where $c_1$ and $c_2$ are functions of $\epsilon$ and are given by Eq.~(\ref{gamma}).
For the even-parity eigenstates we use $\psi_{\rm MAF}^\prime(z=0) = 0$; if we write the first line of Eq.~(\ref{eq: DWP MAF final wavefunction}) in the main text as
\begin{equation}
u(z) \left(c_1 Ai \left[(\frac{3}{2}w_1(z)^{\frac{2}{3}} \right] + c_2 Bi \left[(\frac{3}{2}w_1(z)^{\frac{2}{3}} \right]\right)
 \label{app_mfa8}
 \end{equation}
 (so this merely defines $u(z)$), then one has to find the values of $\epsilon$ that solve
 \begin{eqnarray}
& -{u^\prime(0) \over u(0)} \left(c_1 Ai \left[(\frac{3}{2}w_1(0)^{\frac{2}{3}} \right] + c_2 Bi \left[(\frac{3}{2}w_1(0)^{\frac{2}{3}} \right]\right) = &\ \ \ \nonumber \\
\nonumber \\
& {w_1^\prime(0) \over w_1(0)^{1/3}} \left(c_1 Ai \left[(\frac{3}{2}w_1(0)^{\frac{2}{3}} \right] + c_2 Bi \left[(\frac{3}{2}w_1(0)^{\frac{2}{3}} \right]\right).&(even) \nonumber \\
 \label{app_mfa9}
 \end{eqnarray}
 
\section{Matrix Mechanics} \label{sec:appendix_mat} 

For completeness we provide a quick summary of the equations solved with a matrix diagonalization routine.
Following Ref.~\onlinecite{Matrix_mechanics} (see also Ref.~\onlinecite{Double-potential-well}), we start with the TISE,
\begin{equation}
-{\hbar^2 \over 2 m}{d^2\psi(x) \over dx^2} + V(x) \psi(x) = E\psi(x),
\label{tise}
\end{equation}
with the potential embedded in an infinite square well potential of width $x_c$, and spanning the range $0 < x < x_c$. Unlike Refs.~\onlinecite{Matrix_mechanics, Double-potential-well}
we use the
natural length scale in the problem, which for both the SHO and the DWP, is $x_{\rm HO} \equiv \sqrt{\hbar \over m\omega}$. Therefore, using $z \equiv x/x_{\rm HO}$, and centering the potential of 
interest in the infinite square well, Eq.~(\ref{tise}) becomes
\begin{eqnarray}
-{d^2\psi(z) \over dz^2} + \left(z - {z_c \over 2}\right)^2 \psi(z) = \epsilon \psi(z),   & & {\rm (SHO)} \label{dim_sho}\\ 
-{d^2\psi(z) \over dz^2} + \left(|z - {z_c \over 2}| - z_0\right)^2 \psi(z) = \epsilon \psi(z),  &  & {\rm (DWP)}\ \ \ \  \ \  
\label{dim_dwp}
\end{eqnarray}
with $\epsilon \equiv E/(\hbar \omega/2)$, $z_c \equiv x_c/x_{\rm HO}$, and $z_0 \equiv x_0/x_{\rm HO}$ for the DWP.
Now writing the wave function as a linear superposition of the eigenstates of the (empty) infinite square well,
\begin{equation}
\psi(z) = \sum_{m=1}^{N_{max}} c_m \sqrt{2 \over z_c} \sin{({m\pi z \over z_c})},
\label{expansion}
\end{equation}
with a cutoff $N_{max}$ imposed to make it manageable on the computer, we can substitute into Eqs.~(\ref{dim_sho}) and (\ref{dim_dwp}), and take the inner product with one of the basis states.
The result is the matrix equation
\begin{equation}
\sum_{m=1}^{N_{max}} \left[H^{(0)}_{nm} + V^J_{nm}\right]c_m = \epsilon c_n,
\label{mat}
\end{equation}
where $J \equiv SHO$ or $J \equiv DWP$ for the simple harmonic oscillator or double-well potential, respectively, and
\begin{equation}
H^{(0)}_{nm} = \delta_{nm} \left({n \pi \over z_c}\right)^2.
\label{kin_nm}
\end{equation}
Working out the simple integrals, we obtain
\begin{widetext}
\begin{eqnarray}
V^{\rm SHO}_{nm} &=& 2z_c^2\left(\cos{\left[(n-m) {\pi \over 2}\right]} K_2(n-m) -  \cos{\left[(n+m {\pi \over 2}\right]} K_2(n+m)\right), \ \ \ {\rm and}\\ \label{pot_sho}
V^{\rm DWP}_{nm} &=& 2z_c^2\left(\cos{\left[(n-m) {\pi \over 2}\right]} K(n-m) -  \cos{\left[(n+m {\pi \over 2}\right]} K(n+m)\right), \\ \label{pot_dwp}
\label{pot_nm}
\end{eqnarray}
\end{widetext}
where 
\begin{equation}
K_\ell(n) \equiv \int_0^{1/2}dx x^\ell \cos{(n\pi x)},
\label{akkj_def}
\end{equation}
and for $V^{\rm DWP}_{nm}$,
\begin{equation}
K(n) \equiv  K_2(n) - 2 {z_0 \over z_c} K_1(n) + \left({z_o \over z_c}\right)^2 K_0(n).
\label{akk_def}
\end{equation}
The integrals are elementary; for $n \ne 0$, we have
\begin{eqnarray}
K_0(n) &=& {1 \over n\pi} \sin{({n \pi \over 2})},\nonumber \\
K_1(n) &=& {1 \over 2}K_0(n) - {1 \over (n\pi)^2}\left(1 - \cos{({n \pi \over 2})}\right),\nonumber \\
K_2(n) &=& \left({1 \over 4} - {2 \over   (n\pi)^2}\right)K_0(n) + {1 \over (n\pi)^2} \cos{({n \pi \over 2})},\nonumber \\
\label{akkdef}
\end{eqnarray}
while $K_0(0) = 1/2$, $K_1(0) = 1/8$, and $K_2(0) = 1/24$. Equation~(\ref{mat}) is solved as an eigenvalue equation. Convergence to 5 or 6 significant digits is readily obtained with $z_c \approx 10$ for
the SHO, and $z_c \approx 10 z_0$ for the DWP. The matrix size required is 100 - 200.

\section{Tables} \label{sec:appendix_tables}

In connection with Fig.~\ref{fig: DWP first 4 energies}, we provide the actual eigenvalues determined for the DWP, first using the WKB approximation, 
Table~\ref{tab: DWP WKB Numeric eigvals} and then the MAF approximation, Table~\ref{tab: DWP MAF Numeric eigvals}, followed by the exact eigenvalues, Table~\ref{tab: DWP exact eigvals}
determined numerically through the matrix diagonalization method described in Sec~(\ref{sec:appendix_mat}).
\begin{table}[ht]
    \caption{The first nine WKB dimensionless energies of the DWP for different values of the central barrier height $z_0^2$. These are obtained by numerically solving for the roots of Eq. (\ref{eq: DWP WKB eigenvalue equations}). As $z_0^2$ increases, the difference between energies within each pair of even/odd eigenstates states decreases. This shows a tendency towards degeneracy as the system behaves more like two independent harmonic oscillators.}
    \centering
    \begin{tabular}{|c|c|c|c|}
    \hline
    \ \ n \ \ & $\epsilon$ ($z_0^2=4$) & $\epsilon$ ($z_0^2=9$) & $\epsilon$ ($z_0^2=16$)\\
    \hline
    0 & 0.949292352 & 0.9995628650 & 0.9999994978\\
    1 & 1.039081813 & 1.0003833398 & 1.0000004687\\
    2 & 2.525729513 & 2.9922422377 & 2.9999839988\\
    3 & 3.240818000 & 3.0064271576 & 3.0000146921\\
    4 & N/A & 4.9393381538 & 4.9997693489\\
    5 & N/A & 5.045672658 & 5.000206784\\
    6 & N/A & 6.683940205 & 6.997945408\\
    7 & N/A & 7.175058193 & 7.001777909\\
    8 & N/A & N/A & 8.987266055\\
    \hline
    \end{tabular} \label{tab: DWP WKB Numeric eigvals}
\end{table}

\begin{table}[ht]
    \caption{The first nine MAF dimensionless energies of the DWP for different values of the central barrier height $z_0^2$. These are found by numerically solving for the roots of 
    Eqs.~(\ref{app_mfa9}) and (\ref{app_mfa7}) for even and odd states, respectively. As $z_0^2$ increases, the even/odd states begin to form degenerate pairs with the same energies as was 
    found for the SHO using the MAF method (see Eq.~(\ref{eq: general eigenvalue condition even}) and (\ref{eq: general eigenvalue condition odd}) with the results
    listed in Eq.~(\ref{eq: SHO MAF energies})).}
    \centering
    \begin{tabular}{|c|c|c|c|}
    \hline
   \ \ n \ \ & $\epsilon$ ($z_0^2=4$) & $\epsilon$ ($z_0^2=9$) & $\epsilon$ ($z_0^2=16$)\\
    \hline
    0 & 1.071610296 & 1.120192311 & 1.120662927\\
    1 & 1.155741201 & 1.121065250 & 1.120664017\\
    2 & 2.773368802 & 3.027036049 & 3.034682046\\
    3 & 3.257807700 & 3.040797751 & 3.034713355\\
    4 & N/A & 4.967890267 & 5.02277939\\
    5 & N/A & 5.06300151 & 5.02321277\\
    6 & N/A & 6.81522412 & 7.01379347\\
    7 & N/A & 7.16620891 & 7.01749697\\
    8 & N/A & 8.67856137 & 9.00047532\\
    \hline
    \end{tabular} \label{tab: DWP MAF Numeric eigvals}
\end{table}

\begin{table}[htb]
    \caption{The first 9 exact energies of the DWP for different values of the central barrier height $z_0^2$. These are solved by finding the eigenvalues of the matrix given by Eq.~(\ref{mat})
    from Sec.~(\ref{sec:appendix_mat}).
    Note that all these eigenvalues are converged, meaning that neither increasing the matrix size ($N_{max}$) nor the well width ($z_c$) changes the value of these energies to the number of digits quoted.
    It is clear from the entries for the largest value of $z_0^2$ that
    for large $z_0^2$ these results are approaching those of a doubly degenerate SHO.}
    \centering
    \begin{tabular}{|c|c|c|c|}
    \hline
    n & $\epsilon$ ($z_0^2=4$) & $\epsilon$ ($z_0^2=9$) & $\epsilon$ ($z_0^2=16$)\\
    \hline
    0 & 0.951418841 & 0.999551324 & 0.999999473\\
    1 & 1.035763395 & 1.000390824 & 1.000000491\\
    2 & 2.735035427 & 2.992522095 & 2.999984069\\
    3 & 3.223014022 & 3.006040298 & 3.000014603\\
    4 & 4.670818194 & 4.945524069 & 4.999774618\\
    5 & 5.640887041 & 5.039821328 & 5.000201132\\
    6 & 7.043497484 & 6.798657491 & 6.99802304\\
    7 & 8.266487628 & 7.150150726 & 7.001688794\\
    8 & 9.698852786 & 8.656504400 & 8.987987719\\
    \hline
    \end{tabular} \label{tab: DWP exact eigvals}
\end{table}

\clearpage

\bibliographystyle{IEEEtran}
\bibliography{Bibliography}

\end{document}